\documentstyle[aps]{revtex}
\begin{document}
\newcommand{\pp}[1]{\phantom{#1}}
\newcommand{\be}{\begin{eqnarray}}
\newcommand{\ee}{\end{eqnarray}}

\title{
Einstein-Podolsky-Rosen-Bohm experiment with massive particles as a
test of relativistic center-of-mass position operator
}
\author{Marek Czachor \cite{*}}
\address{
Wydzia{\l}  Fizyki Technicznej i Matematyki Stosowanej\\
 Politechnika Gda\'{n}ska,
ul. Narutowicza 11/12, 80-952 Gda\'{n}sk, Poland
}
\maketitle
\begin{abstract}
The nonrelativistic singlet state average 
$\langle \psi|{\vec a}\cdot\vec \sigma\otimes  {\vec
b}\cdot\vec \sigma|\psi\rangle =-\vec a\cdot\vec b $ 
can be relativistically
generalized if one defines spin {\it via\/} the relativistic
center-of-mass operator.
The relativistic correction is quadratic in $v/c$ and can be
measured in Einstein-Podolsky-Rosen-Bohm-type experiments with
massive spin-1/2 particles. 
A deviation from the nonrelativistic formula would indicate that
for relativistic nonzero-spin particles centers of mass and
charge do not coincide.
\end{abstract}
\vskip1pc

When Einstein, Podolsky and Rosen (EPR) formulated in 1935
their famous paradox \cite{EPR,Bohm}, the main problem they addressed
was an essentially academic question of  completness of the 
quantum theory. Some three decades later Bell \cite{Bell}
derived an inequality which allowed for relating the EPR
{\it Gedankenexperiment\/} to a real experimental situation. 
The photon-pair tests of the Bell inequality performed to
date have ruled out a large class of hidden-variable theories. To
eliminate some of the remaining possibilities one has to violate
the so-called strong Bell inequalities \cite{Home,Kwiat}. For this
reason Fry {\it et
al.\/} \cite{Fry} have recently returned 
to the original idea of Bohm and
Bell and propose to test the strong inequalities by using pairs of
correlated spin-1/2 {\it massive\/} particles. The proposal
involves two $^{199}$Hg atoms, each with nuclear spin
$\frac{1}{2}$, produced in an EPR-Bohm entangled state by
dissociation of dimers of the $^{199}$Hg$_2$ isotopomer using a
spectroscopically selective stimulated Raman process. 

In this Letter I want to show that an EPR-Bohm experiment with
pairs of {\it massive\/} spin-1/2 particles may simultaneously
solve another old (in fact, even older than EPR)
problem of quantum mechanics. 

As is widely known
E.~Schr\"odinger in his 1930  paper \cite{ES} examined
the behavior of the coordinate operator $\bbox x$ 
associated
with Dirac's equation and discovered the oscillatory motion 
he called the {\it Zitterbewegung\/}. The {\it Zitterbewegung\/}
takes place with respect to the {\it center-of-mass\/} position
operator $\bbox x_A$. 
The operator $\bbox x$ is in 
contemporary literature \cite{BB} interpreted as 
the {\it center-of-charge\/} operator,
since it is $\bbox x$ and not $\bbox x_A$ which is used in the minimal
electromagnetic coupling. 
The situation is not typical only of the Dirac equation and is
not associated with the presence of negative energy solutions as
one is sometimes led to believe.
The so-called new Dirac equation generalized by
Mukunda {\it et al.\/}  \cite{Mukunda} admits only
positive-energy solutions but the {\it Zitterbewegung\/} is
present and the associated center-of-mass operator is
algebraically identical to this implied by Schr\"odinger's
analysis of the Dirac equation (cf. the Barut-Zanghi model of
the Dirac electron \cite{BZ}). 
The problem is therefore general and is rooted in the structure
of the Poincar\'e group. 

In what follows I will use a group representation formulation,
elements of which can be found in
the 1965 paper by Fleming \cite{Fleming1}. The group theoretic
approach has the advantage of being applicable to any physical
system whose symmetry group is the Poincar\'e group, or whose
symmetry group contains the Poincar\'e group as a subgroup. The
formulation
is essentially unrelated to the Dirac equation and can be
applied also to hadrons \cite{Mukunda}.

Let us begin with generators of the unitary, infinite
dimensional irreducible representation of the Poincar\'e group
corresponding to a nonzero mass $m$ and spin $j$. Their standard
form is \cite{Ohnuki}
\begin{eqnarray}
\bbox J &=& 
\frac{\hbar}{i}\bbox p\times \frac{\partial}{\partial \bbox p}
+{\bbox s},\\
\bbox K &=& 
\pm\Bigl(
|p_0|\frac{\hbar}{i}\frac{\partial}{\partial \bbox p}
- \frac{\bbox p\times{\bbox s}}{mc+|p_0|}\Bigr),\label{46}\\
\bbox P&=& \bbox p,\\
P_0&=&p_0=\pm \sqrt{\bbox p^2 + m^2c^2}.
\end{eqnarray}
Here {\bbox s} denotes finite dimensional angular 
momentum matrices corresponding to
 the $(2j+1)$-dimensional representation 
$D^j$ of the rotation group.

The center-of-mass position operator which generalizes to any
representation the operator $\bbox x_A$ of Schr\"odinger is
\begin{eqnarray}
{\bbox Q} &=&-\frac{1}{2}\Bigl(P_0^{-1}{\bbox K} +{\bf
K}P_0^{-1}\Bigr) \label{Q}\\
&=&i\hbar\frac{\partial}{\partial \bbox p} 
-i\hbar\frac{\bbox p}{2p_0^2}
+ \frac{\bbox p\times{\bbox s}}{|p_0|(mc+|p_0|)}.
\end{eqnarray}
This operator extends naturally also to massless fields.
Jadczyk and Jancewicz \cite{JJ} found an interesting argument
for its uniqueness in the case of the Maxwell field. 
Orbital angular momentum and spin corresponding to $\bbox Q$ are
\cite{Pryce,Fleming1}
\begin{eqnarray}
\bbox L &=& \bbox Q\times\bbox P=
\frac{\hbar}{i}\bbox p\times \frac{\partial}{\partial \bbox p}
+\frac{|p_0|-mc}{|p_0|}\Bigl(
{\bbox s} -(\bbox n\cdot{\bbox s})\bbox n\Bigr),\\
{\bbox S} &=& \bbox J - \bbox L = \frac{mc}{|p_0|}{\bbox s} +
\Bigl(1-\frac{mc}{|p_0|}\Bigr)(\bbox n\cdot{\bbox s})\bbox n=
\sqrt{1-\beta^2}{\bbox s}_\perp+(\bbox n\cdot{\bbox s})\bbox n.
\end{eqnarray}
${\bbox s}_\perp$ denotes the projection of ${\bbox s}$ on the
plane perpendicular to $\bbox p$ and 
$\beta=|\bbox v|/c$, where $\bbox v=c\bbox p/p_0$ 
is a velocity of the particle.
Projection of spin in a direction given by the unit vector
$\bbox a$ commutes with the Hamiltonian $P_0$ and equals
\begin{eqnarray}
\bbox a\cdot{\bbox S}=\Bigl[\frac{mc}{|p_0|}{\bbox a} +
\Bigl(1-\frac{mc}{|p_0|}\Bigr)(\bbox n\cdot{\bbox a})\bbox n
\Bigr]\cdot{\bbox s}=\bbox \alpha(\bbox a,\bbox p)\cdot{\bbox s}.
\end{eqnarray}
The latter equality 
defines the vector $\bbox \alpha(\bbox a,\bbox p)$ whose length
is 
\begin{eqnarray}
|\bbox \alpha(\bbox a,\bbox p)|=
\frac{\sqrt{(\bbox p\cdot\bbox a)^2 +m^2c^2}}
{|p_0|}.
\end{eqnarray}
The eigenvalues of $\bbox a\cdot{\bbox S}$ are therefore
\begin{eqnarray}
\lambda_{a}=j_3\hbar|\bbox \alpha(\bbox a,\bbox p)|
\end{eqnarray}
where 
$j_3=-j,\dots,+j$.
In the infinite momentum/massless limit the eigenvalues of 
spin
in a direction perpendicular to $\bbox p$ vanish, which can be
regarded as a consequence of the Lorentz flattenning of the
moving particle (in these limits $\bbox S=(\bbox n\cdot\bbox
s)\bbox n$). Projection of spin on the momentum direction is
equal to the helicity, i.e.
$\bbox p\cdot\bbox S=\bbox p\cdot\bbox s$ for any $\bbox
p$, and 
 $\bbox S=\bbox s$ in the rest frame ($\bbox p=0$). 
The definition of spin {\it via\/} the relativistic 
center-of-mass operator can be found already in \cite{ES}. Also
Mukunda {\it et al.\/} \cite{Mukunda} noticed that the extended
models of hadrons based on the generalized ``new" Dirac equation
can be correctly interpreted provided one defines spin {\it via\/} the
relativistic center-of-mass operator (the standard ``natural"
choice of $\bbox s$ leads to physical inconsistencies).
Bacry \cite{Bacry} observed that a nonrelativistic limit of
$\bbox x_A$ leads to a correct form of the spin-orbit
interaction in the Pauli equation if one uses potentials
$V(\bbox x_A)$ instead of $V(\bbox x)$ \cite{Kaiser}; 
an analogous effect was
described in \cite{BB1} where the internal angular momentum of
the {\it Zitterbewegung\/} leads to spin with the correct $g=2$
factor. An algebraic curiosity is the fact that the
components of $\bbox S$ satisfy an algebra which is $so(3)$ in
the rest frame and formally
contracts to the Euclidean $e(2)$
in the infinite momentum/massless limit, and thus
provides an interesting alternative 
explanation of the privileged role
played by the Euclidean group in the theory of massless fields
\cite{ja,Kim}.

In spite of all these facts suggestng
that both $\bbox Q$ and $\bbox S$ are
natural candidates for physical observables no experimental tests
distinguishing them from other definitions of position and spin
have been proposed so far. 

Consider now  two spin-1/2 particles in a singlet state (total
helicity equals zero) and propagating in the same direction with
identical momenta $\bbox p$ (more precisely one should take wave
packets in momentum space, but for simplicity assume that they
are sufficiently well localized around momenta $\bbox p$, so
that we can approximate them by plane waves):
\begin{eqnarray}
|\psi\rangle = \frac{1}{\sqrt{2}}\Bigl(
|+1/2,\bbox p\rangle|-1/2,\bbox p\rangle 
-
|-1/2,\bbox p\rangle|+1/2,\bbox p\rangle
\Bigr). \label{st}
\end{eqnarray}
The kets $|\pm1/2,\bbox p\rangle$ form the {\it helicity\/} basis.
Consider the binary operators
$\hat {\bbox a}=\bbox a\cdot\bbox S/|\lambda_a|$,
$\hat {\bbox b}=\bbox b\cdot\bbox S/|\lambda_b|$.
The average of the relativistic EPR-Bohm-Bell operator is 
\begin{eqnarray}
\langle \psi|\hat {\bbox a}\otimes \hat {\bbox
b}|\psi\rangle =
- \frac{
\bbox a\cdot\bbox b- \beta^2 \bbox a_\perp\cdot\bbox b_\perp}
{\sqrt{1+\beta^2\bigl[(\bbox n\cdot\bbox a)^2 - 1\bigr]}
\sqrt{1+\beta^2\bigl[(\bbox n\cdot\bbox b)^2 - 1\bigr]}}\label{ab}
\end{eqnarray}
There are several interesting particular cases of the formula
(\ref{ab}). First, if $\bbox a=\bbox a_\perp$, 
$\bbox b=\bbox b_\perp$
then 
\begin{eqnarray}
\langle \psi|\hat {\bbox a}\otimes \hat {\bbox
b}|\psi\rangle =-\bbox a\cdot\bbox b
\end{eqnarray}
which is the
nonrelativistic result. If $\bbox a\cdot\bbox n\neq 0$,
$\bbox b\cdot\bbox n\neq 0$ then in the ultrarelativistic case
$\beta^2=1$ 
\begin{eqnarray}
\langle \psi|\hat {\bbox a}\otimes \hat {\bbox
b}|\psi\rangle =-\frac{(\bbox a\cdot\bbox n)\,(\bbox b\cdot\bbox n)}
{|\bbox a\cdot\bbox n|\,|\bbox b\cdot\bbox n|}=\pm 1
\end{eqnarray}
independently of the choice of $\bbox a$, $\bbox b$. 
It is easy to intuitively understand this result: In the
ultrarelativistic limit projections of spin in directions
perpendicular to the momentum vanish for both particles and
spins are (anti)parallel to the momentum. 
The most striking case occurs if $\bbox a$ and $\bbox b$ are
perpendicular and the nonrelativistic average is 0. 
Let $\bbox a\cdot\bbox b=0$, $\bbox a\cdot\bbox n=
\bbox b\cdot\bbox n=1/\sqrt{2}$. Then
\begin{eqnarray}
\langle \psi|\hat {\bbox a}\otimes \hat {\bbox
b}|\psi\rangle =
-\frac{\beta^2}{2- \beta^2}.
\end{eqnarray}
This average is 0 in the rest frame ($\beta=0$) and $-1$ for
$\beta =1$. Any observable deviation from 0 in an EPR-Bohm type
experiment would be an
indication that the operators $\bbox S$ and $\bbox Q$ are
physically correct observables and that massive spin-1/2
particles are extended in the sense that centers of mass and
charge do not coincide. The components of the center-of-charge
operator
commute whereas those of $\bbox Q$
do not commute for nonzero spins. This means that spinning
particles cannot
be localized at a point \cite{Kalnay}. 
This interesting property seems
unavoidable and can be proved at both quantum and classical
levels \cite{Mukunda,Zakrzewski}. Its experimental verification 
could not be without
implications for the self-energy and renormalization problems. 

I am grateful to Ryszard Horodecki for suggesting the problem,
Vasant Natarajan for informations concerning experiments, 
and Gerald Kaiser for extensive discussions. 
The paper is a part of the KBN project 2P30B03809.


\begin{references}
\bibitem[*]{*}Electronic address: mczachor@sunrise.pg.gda.pl
\bibitem{EPR}A.~Einstein, B.~Podolsky, and N.~Rosen, Phys. Rev.
{\bf 47}, 777 (1935).
\bibitem{Bohm}D.~Bohm, {\it Quantum Theory\/} (Prentice-Hall,
Englewood Cliffs, N.J., 1951).
\bibitem{Bell}J.~S.~Bell, Physics {\bf 1}, 195 (1964).
\bibitem{Home}D.~Home and F.~Selleri, Riv. Nuovo Cimento {\bf
14}, No~9 (1991).
\bibitem{Kwiat}P.~G.~Kwiat, P.~H.~Eberhard, A.~M.~Steinberg, and
R.~Y.~Chiao, Phys. Rev. A {\bf 49}, 3209 (1994).
\bibitem{Fry}E.~S.~Fry, T.~Walther, and S.~Li, 
Phys. Rev. A {\bf 52}, 4381 (1995).
\bibitem{ES}E.~Schr\"odinger, Sitzungsb. Preuss. Acad. Wiss.
Phys.-Math. Kl. {\bf 24}, 418 (1930); {\it ibid.\/} 
{\bf 3}, 1 (1931).
\bibitem{BB}A.~O.~Barut and A.~J.~Bracken, 
Phys. Rev. D {\bf 23}, 2454 (1981).
\bibitem{Mukunda}N.~Mukunda, H. van Dam, and L.~C.~Biedenharn, 
{\it Relativistic Models of Extended Hadrons Obeying a Mass-Spin
Trajectory Constraint\/}, Lecture Notes in Physics, No 165
(Springer, Berlin, 1982).
\bibitem{BZ}A.~O.~Barut and N. Zanghi,
Phys. Rev. Lett. {\bf 52}, 2009 (1984);
A.~O.~Barut and W.~Thacker, 
Phys. Rev. D {\bf 31}, 1386 (1985);
W.~A.~Rodriguez Jr., J.~Vaz Jr., E.~Recami, and G.~Salesi, Phys.
Lett. B {\bf 318}, 623 (1993); 
J.~Vaz Jr., W.~A.~Rodriguez Jr., {\it ibid.\/} {\bf 319}, 203
(1993); J.~Vaz Jr., {\it ibid.\/} {\bf 344}, 149 (1995). 
\bibitem{Fleming1}G.~N.~Fleming, Phys. Rev. {\bf 137}, B188
(1965); {\it ibid.\/} {\bf 139}, B963 (1965)
\bibitem{Ohnuki}Y.~Ohnuki, {\it Unitary Representations of the
Poincar\'e Group and Relativistic Wave Equations\/} (World
Scientific, Singapore, 1988).
\bibitem{JJ}A.~Z.~Jadczyk and B.~Jancewicz, 
Bull. Acad. Polon.
Sci. {\bf 21}, 477 (1973).
\bibitem{Pryce}M. H. L. Pryce, 
Proc. Roy. Soc. (London) {\bf
A195}, 62 (1948).
\bibitem{Bacry}H. Bacry, 
Ann. Inst. Henri Poincar\'e {\bf
49}, 245 (1988).
\bibitem{Kaiser}A detailed analysis of the Galilean limit of the
Poincar\'e group and its relation to the center-of-mass position
operator can be found in G.~Kaiser, {\it Quantum Physics, 
Relativity, and
Complex Spacetime: Towards a New Synthesis\/}
(North-Holland, Amsterdam, 1990).
\bibitem{BB1}A.~O.~Barut and A.~J.~Bracken, 
Phys. Rev. D {\bf 24}, 3333 (1981).
\bibitem{ja}M.~Czachor and A.~Posiewnik, ``What happens to spin
during the $SO(3)\to SE(2)$ contraction?", preprint 
quant-ph/9501017.
\bibitem{Kim}Typically the Euclidean structure of massless
fields is interpreted in the language of finite dimensional and
nonunitary representations. It can be shown that the $SE(2)$
structure is associated with a cyllindrical geometry: The
cyllinder is parallel to momentum and the group contains
rotations around and translations along the cyllinder. The
translations are gauge transformations of a vector potential.
The relativistic spin algebra discussed in this Letter provides an
explanation of the contraction in terms of the Lorentz
flattenning of an extended particle and the corresponding
deformation of the spin algebra. The analogy between the two
approaches is however even deeper: The noncommuting position
operator allows for localizations of the Maxwell fields on
curves \cite{JJ} and the classical phase space picture derived
in \cite{Zakrzewski} leads to world-tubes rather than
world-lines. An analogous phenomenon is found in the twistor
formulation where instead of world-lines the massless fields are
interepreted in terms of the Robinson
congruence of twisting null world-lines
\cite{PR}. The cyllindrical approach is described in 
Y. S. Kim and M. E. Noz, {Phys. Rev.} D {\bf 15}, 335 (1977);
Y. S. Kim, {Phys. Rev. Lett.} {\bf 63}, 348-351 (1989).
D. Han, Y. S. Kim, and D. Son, {Am. J. Phys.} {\bf 54}, 818
(1986);D. Han, Y. S. Kim, and D. Son, {Phys. Lett.} {\bf
131B}, 327 (1983);
D. Han and Y. S. Kim, {Am. J. Phys.} {\bf 49}, 348 (1981);
J. J. van der Bij, H. van Dam, and Y. J. Ng, {Physica} {\bf 116A},
307 (1982);
D. Han, Y. S. Kim, and D. Son, {Phys. Rev.} D {\bf 26}, 3717
(1982); 
Y. S. Kim and M. E. Noz, {Theory and Applications of the Poincar\'e
Group} (Reidel, Dordrecht, 1986);
Y. S. Kim and E. P. Wigner, {J. Math. Phys.} {\bf 28}, 1175 (1987);
{\it ibid.\/} {\bf 31}, 55 (1990).
\bibitem{Zakrzewski}
S.~Zakrzewski, ``Extended phase space for a spinning
particle", preprint hep-th/9412100.
\bibitem{PR}R.~Penrose and W.~Rindler, {\it Spinors and
Space-Time\/}, vol.~2 (Cambridge University Press, 1986);
A.~Bette, 
J. Math. Phys. {\bf 25}, 2456 (1984); A.~Bette,
Rep. Math. Phys. {\bf 28}, 133 (1989).
\bibitem{Kalnay}Actually, many different position operators have
been proposed so far. A review of the problem can be found in 
A.~J.~Kalnay, ``The Localization Problem", in {\it Studies in
the Foundation, Methodology and Philosophy of Sciences\/},
vol.~4, edited by M.~Bunge (Springer, 1971), 
A.~O.~Barut and R.~R\c{a}czka, {\it Theory of Group
Representations and Applications\/} (Polish Scientific
Publishers, Warszawa, 1980), and H.~Bacry, {\it Localizability
and Space in Quantum Physics\/} (Springer, Berlin, 1988). 
A detailed study of commuting
operators in the context of massless fields can be found in 
E.~Angelopulos, F.~Bayen, and M.~Flato, Phys. Scr. {\bf 9}, 173
(1974). 
\end{references}
\end{document}